\documentclass[12pt,a4paper]{article}
\usepackage{amsmath,amssymb}

\usepackage[pdftex]{graphicx}
\usepackage{cite}

\allowdisplaybreaks[3]

\begin{document}

\begin{titlepage}

\begin{flushright}
KEK-TH-2223, 
\end{flushright}

\vspace{0.5em}

\begin{center}
{\Large{\textbf {Curvature Perturbations and \notag\\
\vspace{1em}
 Anomaly explain Dark Energy}}}
\end{center}

\begin{center}

Yoshihisa \textsc{Kitazawa}$^{1),2)}$
\footnote{E-mail address: kitazawa@post.kek.jp} 
\end{center}

\begin{center}
$^{1)}$
\textit{KEK Theory Center, Tsukuba, Ibaraki 305-0801, Japan}\\
$^{2)}$
\textit{Department of Particle and Nuclear Physics}\\
\textit{The Graduate University for Advanced Studies (Sokendai)}\\
\textit{Tsukuba, Ibaraki 305-0801, Japan}\end{center}

\begin{abstract}

We investigate the history of dark energy to explain the present magnitude.
We assume the dark energy is the residual cosmological constant.
The most important channel in the reheating process is the gluon pair productions by QCD trace anomaly.
We argue dark energy decays rapidly   by  gluon pair emissions during the reheating and
 after the big bang.
  The reheating temperature is determined  by the decay width
 of dark energy $\Gamma$ and the Planck mass $M_p$
 as  $\sqrt{M_P\Gamma} \sim 10^6GeV$.
It is the consequence of Friedmann's equation and 
an equilibrium condition $\Gamma\sim H$.
As the Universe cools below the hadronic scale,
dark energy density is almost frozen. Nevertheless the  dark energy further decreases by emitting two photons. 
We have estimated the current decay rate of dark energy from the  QED trace anomaly.
The consistent solution of Friedmann equation is in  an excellent agreement
with the observations. The suppression factor of dark energy scale is the product of  fine structure constant $\alpha$ and curvature perturbation $P$ as $10^{-30}=(\alpha^2P/4\pi)^2$ .
We argue the conformal symmetry breaking in the both UV and IR are necessary unless
dark energy is subtracted. 
We also investigated lepto-genesis by adding massive right handed neutrinos.
The realistic  lepto-genesis takes place during reheating process.

\end{abstract}

\vspace{\fill}

June 2021

\end{titlepage}

\section{Introduction}
\setcounter{equation}{0}
Cosmological constant problem has been recognized as one of the greatest
 riddles in physics \cite{WS} . If it is non-vanishing, it could be at one of the fundamental scales.
 The most natural scale is the Planck scale $M_P$. However the present value turns out to be much smaller than that by the factor $10^{30}$. The dark energy density   $(10^{-3}eV)^4$ is of familiar magnitude for us. 
 It might be related to biology . Although  such a possibility cannot be ruled out, there might be a simpler solution \cite{Polyakov}\cite{Jackiw}\cite{TW}. In fact, the universe may have started with de Sitter expansion\cite{KKN}.

The greatest mystery here is  the identity of inflaton and its relation to dark energy.
We have shown that the linear inflaton potential $V(f)=1-2\gamma f$ is generated at the one-loop level \cite{KKM}, 
\begin{align}
\int \sqrt{-g}d^4x \Big[\frac{1}{\kappa^2}(R-6H^2V(f)-{2\gamma}\partial_\mu f\partial^\mu f)\Big].
\label{action2}\end{align}
where $\kappa^2=16\pi G_N$.
This inflation theory  is dual to the following action  on shell and the classical solutions are 
$f_c=\log a_c=Ht$.
\begin{align}
\int d^4x \Big[\frac{1}{\kappa^2}(a^2\tilde{R}-6H^2a^{4}(1-2\gamma\log a)+6\partial_\mu a\partial^\mu a)\Big].
\label{action2p}\end{align}
$a$ denotes the conformal mode of the metric and (\ref{action2p}) is the Einstein action plus IR logarithmic correction.
The slow-roll parameter $\epsilon=(V'/V)^2\kappa^2$ agrees with the  one loop IR logarithmic correction to $V=1-2\gamma\log(a)$.
Here $\epsilon =\gamma=(3/8\pi)(H^2\kappa^2/ 4\pi )$. 

This coincidence suggests de Sitter duality. Namely IR log effects in quantum gravity
broke de Sitter symmetry. This effect corresponds to inflaton potential in classical gravity.  Thus, the inflation theory with the particular potential is equivalent to the quantum gravity with IR log effects. The duality explained here holds at weak coupling $g \log Ht < 1$.
We need to sum up IR logs to all orders if we are interested in the past or future of the Universe. For that purpose,
we employ Fokker Planck equation. There is an indication the resummation of IR  logs may generate non-perturbative potential
for degenerate vacua. We suspect dark energy may be such a problem.
We thus interpret dark energy as the residual
cosmological constant. 

Cosmic magnetic fields of $O(10^{-6}\textmd{Gauss})$ are observed in the galaxies and in the cluster of galaxies.
The magnetic energy density almost coincides with that of CMB. It is too small in comparison to dark energy now.
Although they might be of a primordial origin,
the classical conformal invariance of gauge theory prohibits primordial magnetic field.
Quantum effects (anomalies) may be relevant to this problem.
The magneto-genesis requires an out-of-thermal equilibrium condition and a macroscopic parity violation. These conditions could be naturally satisfied by the EW and QCD phase transitions.
Although this is an interesting problem to explore further, 
it is beyond the scope of this paper.


 Our basic strategy is to solve Friedmann's equation 
\begin{align}
3H^2M_P^2=\rho
\label{FEQ}
\end{align}
 with the equilibrium condition $\Gamma\sim H$.

In (\ref{FEQ}), $\rho$ denotes the energy density. The big bang begins as the thermalized radiation
$\rho \sim T^4$ where $T$ is
the reheating temperature.
On the other hand,  dark energy dominates $\rho$ now $(70\%)$.
Due to the $\omega- h-h$ couplings in  the anomaly where $h$ denotes a gluon, the dark energy  decays by emitting a pair of gluons. 
During inflation,
 every local patch close enough to horizon goes out and then comes back. This process must continue for a sufficient duration. Since there is a flat direction in the potential, super-horizon mode is frozen. It begins to oscillate when it reenters the horizon.  So near the horizon, particle production
takes place.
The energy flow is associated with this procession.
With supply of energy, dark energy turns into thermal  gluons.
The condition $H\sim\Gamma$ determines the beginning of radiation dominated
Universe. By the first law, the entropy flows out during the inflation
and turns into hot radiation in the reheating process. 
We argue curvature perturbations play the central role not only during reheating process but into the big bang.
In order to derive big bang from inflation, reheating must thermalize the Universe. If so,
(\ref{FEQ}) determines the reheating temperature  $\sqrt{M_p\Gamma}$ 
as shown in (\ref{SCFE}).
 Since the conformal mode is a part of space time metric, $\sqrt{-g}=\exp(4\omega)$, the reheating temperature is low $O(\sqrt{HM_P})\sim 10^6GeV$ as it is suppressed by the gravitational coupling.

After the reheating, the most of energy of inflaton is transformed into radiation energy.
The big bang begun with the radiation dominated thermal states and it is a very successful theory.
To our surprise the Universe resumed accelerated expansion when $z<0.6$.
The energy contents of current Universe is $\text{matter: dark energy}=0.3: 0.7$.
 At the cosmic scale, the Universe started accelerated expansion again now just like at the beginning. 
In order to solve
the cosmological constant problem, we need to explain not only its magnitude
of $(10^{-3}eV)^4$ but also why it dominates now. In other words we need to explain the history
of the universe.
In this paper we present our idea on the history of dark energy.
Inflation  and reheating set important stages in our story.

In our view dark energy is the left over cosmological constant. It can decay through QCD or QED conformal anomaly
by emitting a pair of gluons or photons. So it continues to decay throughout the cosmological history.
In fact it once was the most unstable form of energy (in comparison to radiation and matter). That is why we did not notice it until hadronic decay channels are closed due to QCD phase transition.
It still has the weak instability in the QED sector. It decays by emitting two photons through QED conformal anomaly.
We can estimate the 
QED decay width $\Gamma_{QED}$.
Surprisingly the equilibrium condition $\Gamma_{QED}\sim H$ reproduces the temperature  when dark energy dominance begins  $\sqrt{M_P\Gamma_{QED}}\sim 10^{-3}eV$.

The curvature perturbation and conformal mode scales identically with respect to the wave number $k$ \cite{YK}. They are physically the same.
So our scenario works in generic inflation/quintessence theories \cite{InfAG}\cite{ksato}\cite{InfLD}\cite{InfPS}\cite{RP}\cite{CDS}.

The possible connection between anomaly and inflation has been suggested in the seminal work\cite{Infst}.  Our new contribution is that QCD
and QED trace anomaly play a crucial role in the reheating process.

The contents of this paper are as follows.
In this section, we introduced our strategy toward
 a resolution of the cosmological constant problem. In
 section 2, we investigate the reheating process 
:the transition from inflation to big bang in details. 
In section 3, we investigate the decay of dark energy due to conformal anomaly.
In section 4, We point out that the dark energy becomes very stable after QCD transition.
That is the reason why it dominates the energy contents of the Universe now. We further find
the magnitude of dark energy in  an excellent agreement
with the observations.
We conclude in section 5 with discussions.

\section{Active dark energy} 
\setcounter{equation}{0}

We first list the two milestones of the Universe. 
\begin{align}
&\text{The reduced Planck scale: }M_P=\sqrt{1\over 8\pi G }\sim 2.4 \times 10^{18}(GeV).\end{align}
We here introduce the power spectrum of the 
curvature perturbation $P$
\begin{align}
P= R^2={H^2\over (2\pi)^2 2M_P^2\epsilon}\sim 2 \times 10^{-9}.\label{CP}
\end{align}
$\epsilon $ is a slow roll parameter which must be small for a consistent inflation theory. $(H/M_P)^2$ gives the strength of the primordial gravitational wave. The curvature perturbation cited
in (\ref{CP}) is well known as  it has seeded the large scale structure of the Universe. It is the value at the horizon in CMB measurements. 
It is appropriate data for our study of dark energy.  We denote the magnitude of the wave number by $k=|\vec{k}|$. The scale independence of $P$ implies $k$ independence of it. $k$ dependence can be safely ignored for $z<1000$ as its scale independence holds well\cite{Planck18}.
We show that it accounts for the notorious suppression factor $10^{-120}$ of the dark energy density. 

Although it is customary to assume that the e-folding number is $50<N<60$,
we reexamine it  briefly as it is relevant to the history of dark energy.

It has been estimated as\cite{LL}
\begin{align}
N=63.3+{1\over4}	\log \epsilon +{1\over 4} \log {V_{hor}\over \rho_{end}}
+{1\over 12}\log{\rho_{reh}\over \rho_{end}}.
\end{align}
 where $\rho_{hor},\rho_{end},\rho_{reh}$ denote the energy density at the horizon exit, 
the end of inflation and the reheating era respectively. $V_{hor}$ is the potential at the horizon exit.
We investigate the mechanism in which  the conformal mode plays an important role. Thus 
we investigate a weakly coupled theory with a relatively small reheating temperature $T_R$.
This fact implies the last term is very important. 
Nevertheless, we have no firm ground to assume that $N$ is much smaller than $50$.

The conformal mode of the metric couples to the standard model particles via anomaly.
It also couples to heavy fields such as the right-handed neutrino 
and the dark matter through the mass terms.
The gravitational coupling gives rise to the suppression factor $H^2/M_P^2\sim \epsilon 10^{-7}$. The only upper limit is known for one of the slow roll parameter $\epsilon<1/200$. The $\beta$ function in QCD is the largest standard model coupling of $O(1)$.
They play an important role in this paper until the Universe cools down to the hadronic scale.
It is because gluons become massive.
In this respect, QED is most relevant to the dark energy problem since a photon remains massless. 
The right-handed neutrino and the dark matter may contribute significantly to
the reheating process. However they are relevant only at very high energy scale
since they cannot be produced otherwise energetically.

It is important to understand the basic mechanism behind reheating\cite{ABYRM}.
Let us consider the following equation of motion
\begin{align}
\ddot{\chi}+q\omega(t)\chi=0.
\label{eqmot}
\end{align}
Here $\chi$ denotes a particle such as a gluon which couples
 to time dependent curvature perturbation.
$\omega(t)$ denotes the curvature perturbation or conformal background. 
The second term in (\ref{eqmot}) is classically absent on FRW. It arises due to quantum effect: conformal anomaly as in (\ref{anomalous act}).
The coupling $q$ is assumed to be
small.

As is often the case, the driving force is the harmonic oscillators with time dependent frequency.
Particle production takes place when the   adiabaticity is violated.
Although $\omega(k) $ is constant outside the horizon $kt<1$, reheating takes place when it  reenters the horizon $kt>1$.
Since we are interested in the behavior of dark energy near the horizon,
we adopt the following approximation $<\omega^2>_{out} \sim <\omega^2>_{horizon}$.

The behavior of  $<\omega^2>$ has been observed in the scalar power spectrum in CMB. 
Although it is not flat in general, it is approximately so for small angular momenta of horizon scale.
We focus on those modes just  got inside the horizon.

In dark energy case, we find  the  self-consistent weak coupling approximation  works well. It is especially true now since the effective coupling is the fine structure constant $\alpha_e$.
After the inflation, the Universe is reheated due to the particle creation 
in the time dependent background. In this process more entropy is generated
in the form of hot radiation.
 Total entropy of CMB is $10^{45}$ while 
de Sitter entropy is $10^{120}$. 
After the reheating, the radiation dominated thermal equilibrium is realized.

We argue that the pair creation of gluons  dominate thermalization process toward big bang due to the following 
QCD conformal anomaly term 
\begin{align}
\int d^4x({1\over \alpha_s }-2\beta \omega ){1\over 8\pi}tr[F_{\mu\nu}F^{\mu\nu}].
\label{anomalous act}
\end{align}
where
 $\beta=(33-2N_f)/(12\pi)$ and $\alpha_S=g^2/4\pi$
is the coupling of the strong interaction \cite{GW,HP}.

If we replace  the conformal mode $\omega$ by  the neutral $\pi^0$,
we obtain an action for chiral anomaly.

\begin{align}
\int d^4x{\alpha_e\over 2\pi f}\pi^0F_{\mu\nu}\tilde{F}^{\mu\nu}.
\label{anomalousp}
\end{align}
where $\tilde{F}^{\mu\nu}=\pm (1/2) \epsilon ^{\mu\nu\alpha\beta}F_{\alpha\beta}$.

It makes the prediction for $\pi_0$ decay width $\Gamma_{\pi^0}=\alpha_e^2m^3/64\pi^3f^2$.
We evaluate $\Gamma_{\omega}$ for the conformal mode $\omega$ from (\ref{anomalous act})
in an analogous way.
We assume the following correspondences.
\begin{align}
\pi_0:{\textrm pion} &\leftrightarrow \omega :{\textrm  inflaton},\notag\\
e^{i\pi_0/f}  &\leftrightarrow e^{\kappa\omega}.
\end{align}
The inflaton is represented by $\omega$ or $e^{\kappa\omega}$.

In the explanation of the decay mechanism, we draw the analogy with the neutral pion decay. 
Since the pion is pseudo scalar , the relevant decay vertex is $W\sim\pi_0F\tilde{F}$. On the other hand,
the inflaton is Lorentz scalar as we identify it with conformal mode $\omega$ . Although the vertex changes as $W\sim\omega F^2$, the decay width is  $\Gamma\sim O(T)$ in the both cases on the dimensional grounds.
The important point is that anomalies provide the universal decay mechanism even if they are different types.

After reentering the horizon, the conformal mode (curvature perturbation) oscillates with time.
  The 3 point vertex $\omega-h-h$ after the Fourier transformation is
\begin{align}i&\int dt \int d^3x
e^{i(k_1-k_2-k_3)t}e^{i(\vec{k}_2+\vec{k}_3)\cdot \vec{x}}\omega({k}_1)
tr[F^{\mu\nu}(\vec{k}_2)F_{\mu\nu}(\vec{k}_3)]
\notag\\
=&(2\pi)^4\delta (k_1-k_2-k_3) \delta ^3(\vec{k}_2+\vec{k}_3)  
\omega^*({k}_1)
tr[F_a^{\mu\nu}F^a_{\mu\nu}]
\end{align}
We assume conformal mode oscillates as $\omega(k_1) e^{ik_1t}$ inside the horizon.

This is the primary source of reheating.
The kinematics is identical for a massive particle of $m=T$ decays into
two massless particles. 
The big bang requires the thermalization  of the radiation  while the inflaton decays.
In addition, we have $SU(3)$ color indices.
	 $T$ is the parameter (temperature) of the reheating Universe.
	As it turns out , it is a good strategy to use the temperature, or  Matsubara frequency $T$
instead of the Hubble parameter $H$.
It is because we are looking for thermal equilibrium inside horizon. They must possess the common temperature or  Matsubara frequency $T$.
Furthermore, the decay width of dark energy depends on the temperature linearly.

We investigate a simple reheating process.
 We consider the time-like off-shell  conformal mode $\omega(k_1)$
decays into two on shell gluons $h({k}_2),h({k}_3)$.
To be precise, we assume the following kinematics.
\begin{align}
 {k}_1={k}_2+{k}_3, ~~
{k}_1^2=T^2,~~{k}_2^2=0,~~{k}_3^2=0 .
\end{align}
\begin{align}	
W(\omega(k_1))W^*(\omega(k_1))=
{\beta^2\omega^2(k_1)\over 2}<F^a_{\mu\nu}F_b^{\rho\sigma}><F^b_{\rho\sigma}F_a^{\mu\nu}>
={1\over 2}\beta^2\omega^2(k_1).\end{align}

After the use of this algebraic identity, we obtain
\begin{align}
{2\pi\over 2T^3}\int {d^3k_2\over (2\pi)^3}
&\delta (k_1-k_2-k_3)W(\omega(k_1))W^*(\omega(k_1)),
\notag\\
&={T\alpha_S^2\beta^2\omega^2(k_1)\over
16\pi} \times 8.\label{2pc}
\end{align}

In the final step of the reheating calculation (\ref{2pc}),
it is necessary to evaluate two point function of the conformal mode $\omega^2$ as they supply energy inside horizon
by oscillations.

The FP formalism implies $\omega (t)$ is a stochastic variable which satisfies Langevin equation
\begin{align}
	{\delta \omega (t)\over \delta t}={\sqrt{3}\over 2}X(t),~~<X(t)X(t')>={g\over \xi}H\delta (t-t').
	\label{Lgeq}
\end{align}
where $X(t)$ is the massless mode $\sim \zeta$.
Thus the zero mode performs Brownian motion. 
\begin{align}
	<\omega(t)^2>&= \int^t dt' {3 g\over 4\xi}H,\notag\\
	 {\partial\over \partial N} <\omega(t)^2>&={3g\over 4\xi}.
	\label{Lgsp N}
\end{align}

 de Sitter duality predicts that inflationary  universes                                                                                                                                                                                                                                                                                                                                                                                               appear as the solutions of FP equations\cite{KKM}
 \begin{align}
 {\partial \over \partial N}\log {g\over \xi}= 6\xi .
 \end{align}
In fact we find such solutions for the power potentials labeled by real positive parameters $m$\cite{YK}
\begin{align}
g=\tilde{N}^{m\over 2}, ~~6\xi=-{m+2\over 2\tilde{N}}=n_s-1, ~~\epsilon = {m\over 4\tilde{N}}.
\label{strong}
\end{align}
where $\tilde{N}=N_0-N$ since $g$ becomes small as the e-folding $N$ becomes large.

The  horizon crossing time $t_*$ is
\begin{align}
{ \dot{N}(t_*)e^{N(t_*)}\over k} \sim 1.
\end{align}
Equivarently, $k/e^N(t_*)\sim H$.
The $t_*$ dependence leads to additional $k$ dependence in the 2-point correlation function.
The $k$ dependence is readily recovered from time: $N$ dependence. The Horizon is the only finite physical location in de Sitter space. We consider the effective theory with the renormalization scale of the boundary. It resumes IR logarithms by renormalization group. They reflect the random walk of super horizon modes constantly jolted by horizon exiting modes.

The $k$ dependence of the curvature perturbation and conformal perturbation can be estimated from
(\ref{strong}):
\begin{align}
{\partial \over \partial N}\log({g\over \epsilon})= {\partial \over \partial N}\log {3g\over 4\xi}= n_s-1.
\label{Lgsp1}
\end{align}
After integration over $N$, we recover $k$ dependence
\begin{align}
{g\over \epsilon}\propto{g\over \xi} \propto k^{n_s-1}.
\end{align}

So far we have resumed IR logarithms by FP equation.
The result is the emergence of the slow roll parameters such as $n_s$ in (\ref{Lgsp1}) which
spoil de Sitter symmetry.

In the literature, $\delta N$ formalism is widely used to investigate the curvature perturbation. It underscores the validity of the stochastic picture of the inflation
\cite{Starobinsky1985},\cite{Bond,Stewart,Lyth2005a,Lyth2005b}.
 Let us consider the fluctuation of the curvature perturbation $\zeta$.
\begin{align}
{ \zeta}& = {\delta N} ={H\over \dot{ \varphi}}\delta\varphi .
\label{Lgdn}
\end{align}

We obtain in the super-horizon regime.
\begin{align}
<\zeta^2(t)> &=<  ({H\over \dot{\varphi}})^2\delta\varphi^2>\notag\\
=&<{1\over 2\epsilon M^2_P}\delta\varphi^2>= \int_k^H{dk'\over k'}{H^2\over 8\pi^2\epsilon M^2_P}.\notag\\
{\partial \over \partial N}<\zeta^2(t)>&= {H_*^2\over 8\pi^2\epsilon_* M^2_P}\sim
{1-n_s} .
\label{P123}
\end{align}

 The curvature perturbation $\zeta$ and conformal mode $\omega$ obey analogous                                                                                                                                                                                                                   Langevin type equations (\ref{Lgdn}) and  (\ref{Lgeq}). They belong to the same universality class as
two point functions (\ref{Lgsp1}) and (\ref{P123}) scale in the same way   $k^{1-n_s}$.
The conformal mode is indistinguishable from the curvature perturbation. Our identification $<\omega^2>=DP$  therefore works in generic  quantum gravity with conformal  mode or inflation theory with curvature perturbation. The advantage of conformal mode is  being Lorentz scalar, well understood in covariant field theory. Since we have no dimensionful quantity in our equations
(\ref{Lgsp1}) and (\ref{P123}), they must arise as integration constants. In COBE normalization case, the precise coefficient $D$ needs to be fit with the data. We assume it is $O(1)$ by naturalness.

We have proposed a formalism in which the cosmological constant decays via trace anomaly.
The necessary energy is provided by the inflow of the curvature perturbation.
We argue it is a universal decay mechanism of dark energy.
Although $P$ is constant outside the Horizon, we are interested in its behavior inside Horizon.
In that case, $P$ is no longer constant. For example, CMB exhibits acoustic oscillations inside Horizon.
We concentrate on the region where $P$ is well approximate to be constant.

Anomaly is the unavoidable symmetry breaking mechanism in quantum field theory. 
QCD anomaly is a nice potential which contains 2 strongly coupled standard model fields and 
one curvature perturbation.
The conclusion is that dark energy decay phenomenon   due to anomaly may be universal
as it is unavoidable.
The history of dark energy might be also universal.
It is exciting to investigate them further  from the both theoretical and observational point of view.

The universal result is
\begin{align} 
\Gamma\sim {\pi\over 2}\alpha_S^2 \beta^2 PT \sim (0.1)^310^{-10}T
\sim 10^{-12}T (GeV). \label{F2}
\end{align}
The linear dependence of $\Gamma$ on $T$ is kinematical effect.
Since the above $\Gamma\sim H$ is finite, the dark energy is found to decay.
We thus conclude 
\begin{align}
H\sim \alpha^3R^2T.
\label{F1}
\end{align}
Note that this formula contains the curvature perturbation $R^2$ and conformal anomaly $ \alpha^3$.
Even in the big bang era, the dark energy continues to decay.
However this QCD anomaly instability goes away when the Universe cools down below $100 MeV$
and QCD phase transition has taken place. It is because gluons become massive
and the dark energy cannot provide enough energy to pair produce them.

We still need to check the possible instability against  photon pair production.
As it turns out that is the most important effect on dark energy when
$T<100MeV$.

The necessary condition for quasi stationary state is $\Gamma=H$.
Since $\Gamma   << H$ at the initial stage of reheating, it takes a long time to reach the 
equilibrium.
After big bang, the effect of dark energy disappears relative to radiation and matter as it decays exponentially.

 The Friedmann equation can determine
the temperature of the big bang in a self-consistent way.
\begin{align}
3H^2M_P^2\sim 3\Gamma^2M_P^2=\rho_{\omega}+\rho_{r}\sim 30T^4.
\label{SCFE}\end{align}

where $30T^4$ denotes the energy density of the standard model.
In particular, we consider the $(50,50)$
radiation and dark energy balance point.
\begin{align}
3\Gamma^2M_P^2=3\cdot 10^{12}T^2\sim 30T^4.
\end{align}
The big bang starts  at  $T\sim 10^{6}GeV$ as the radiation dominant thermalized state. 
 $\rho_{\omega}$ decreases rapidly but non-vanishing. We assume it is the predecessor of
dark energy. 

 Needless to say, there could be considerable uncertainty on  our knowledge of $P$ to apply it
in the beginning of big bang. We should be prepared to be surprised in this field.
 We have found the decay width of dark energy rather low $10^{-6}GeV$. Its life time is
 $10^{-12}$ seconds. Its survival probability at  the QCD transition period is estimated as
 $e^{- 10^2}\sim 10^{-50}$.
On the other hand ,  let us compare it to the other stuff 
\begin{align}
 N=\int_{t_i}^{t_f} dt H\sim {1/2}\log ({t_f/t_i}),\rightarrow a(t_f)/a(t_i)\sim10^{10}.
 \end{align}
 
 Let us assume the big bang started with radiation $(50\%)$ and dark energy $(50\%)$ at reheating temperature $10^{6}GeV$.
We find the dark energy decays faster $(1/a^5)$ than the radiation $(1/a^4)$. 

\section{Matured dark energy}
\setcounter{equation}{0}

Since gluons have obtained a mass gap by QCD phase transition, 
we only need to  consider the pair creation of photons
as the total decay width $\Gamma_{QED} $ of dark energy below QCD scale. 

QED conformal anomaly action is
\begin{align}
\int d^4x({1\over \alpha_e } +\beta_e \omega) {1\over 4
    \pi}F_{\mu\nu}F^{\mu\nu},
\end{align}
where
 $\beta_e=Q^2/(3\pi), Q^2=\sum_iq_i^2=8$. We have changed the gauge group from $SU(3)$
 to $U(1)$.
$\alpha_e=e^2/4\pi$ is the fine structure constant.
Here we have taken account of the contribution to $\beta$ from all charged fermionic particles in the standard model.  

It can be obtained from $\Gamma$ (\ref{F2})
by  dropping the color multiplicity $8$ and then replace other quantities by that of $QED$.
\begin{align} \Gamma_{QED}\sim ({1\over 4\pi})^2 {\pi\over 2} \alpha_e^2 \beta_e^2 P T_{DE}\sim 10^{-15}T_{DE} .
\label{TDE}
\end{align}

$H$ decreases with $T_{DE}$. It vanishes if 
$P$ is absent.
Note we have introduced a different energy scale for dark energy $T_{DE } << T$ .
 We assume the estimate of the curvature perturbation $P$  is reliable in this regime 
as it exits the horizon at the beginning of the inflation.
It also has changed little after entering into the horizon as dark energy becomes stable.

We first investigate how the Universe resumed accelerated expansion recently. Our critical solution is such that the Universe resumes the acceleration at $(120,80)$ matter, dark energy density ratio at $z=0.6$. It should have evolved into  $(30:70)$ system  by now $z=0$ \cite{KKM}. 
Below the critical temperature, the Universe expands with accelerated speed.
In the parametrization $H_0=100 \textrm{h km/s/Mpc}$, $h$ is found to be around $0.7$ at $z=0$ as shown in Fig. 1. 

Since  $Mpc=3 \times 10^{22} m=10^{5}\cdot 3\times {10}^{17} m$,
the Hubble parameter is 
$H_0=2\times 10^{-18}{1/s}=(2/3) \times 10^{-26}{1/m}={4/3}\times 10^{-33}eV$.  
The life time of dark-energy  is determined to be $1/6 \times 10^{11}$ years in a good agreement with the current age of the Universe. It is an evidence that our equilibrium assumptions $\Gamma=H$ and $H>H_0$ hold.

It corresponds to $T_{DE}=10^{-18}eV$.
The reheating  temperature $T_c$ is $\sqrt{M_PH_0}=10^{-3}eV$.
It is the energy scale when the domination of  dark energy began.
We have fixes the energy scale of the Universe uniquely.
$H/T_{DE}\sim\alpha^3P\sim 10^{-15}$ follows from  (\ref{TDE}). 
The notorious factor $10^{-30}$ turns out to be decomposable into the product of small parameters as $\alpha^6P^2$.

$e^{-\Gamma_{QED} t}\sim\rho$ is the surviving dark energy density.
In the equilibrium, $\rho\sim e^{-N}\sim 1/a(t)$.
Furthermore $\rho(t)\propto 1/a(t)\propto 1+z$ behavior of dark energy is consistent with the observations when $z>0.6$
as shown in Fig.1.\footnote{Private communication. I thank Takahiko Matsubara for his interest in my work.}
Dark energy behaves like $1+z$ above the critical point and below the hadronic scale.
\begin{align}
{H^2(z)\over H_0^2}=
0.5(1+z)+0.3(1+z)^3+10^{-4}(1+z)^4,~~0.6< z < 3500.
\end{align}
The acceleration of the Universe can be estimated by 
$2\gamma H^2=(1+z){\partial \over \partial z}H^2$.
The critical point exists at $z=0.6 $ since $\gamma=1$ at this point.

We expect our theory needs to be refined when $z<0.6$  as the Universe resumed the accelerated expansion. 
We have investigated $0<z<0.6$ region in \cite{KKM} and the modification is
included in Fig.1  as the log term in the following formula.
\begin{align}
{H^2\over H_0^2}=0.3(1+z)^3
+0.7\log(e+\log(1+z)) \theta(0.6-z).
\end{align}
The log term is necessary to reproduce  the current energy contents $(30:70)$. 
$\gamma=1$ is maintained at $z=0.6$.

We find a startling  result  as it reproduces observed magnitude of dark energy at the critical point
by the Friedmann equation and conformal anomaly. 
Furthermore  our reconstruction of its history is consistent with the observation.
The magnitude of current dark energy is  related to the life time of it due to the condition for equilibrium $\Gamma=H$.
The question of why now and a history of dark energy are intermingled as expected.
The history has  assured us dark energy  is consistent with general relativity as it obeys Friedmann equation. The recent dominance may be 
due to the fact that dark energy  becomes the most stable as the Universe cools down.

It should be possible to consolidate detailed history of dark energy of the Universe by further investigations.
We are optimistic with respect to the validity of our proposal  why dark energy
dominates now. It is because dark energy decays more rapidly in comparison to matter energy after the reheating and big bang.
Furthermore it becomes almost constant in the hadronic phase.
Nevertheless two photon production process appears to reduce dark energy slowly
since Friedmann equation is maintained.
By imposing equilibrium condition, we can probe the dark energy when it
emerges as the dominant energy contents of the Universe.

Fig.1 and that of \cite{KKM} coincide at $0<z<0.6$
when the expansion of the Universe is accelerating.
In contrast, they differ when $z>0.6$. Fig.1 includes the effect of reheating as a novel result.
It shows the stability of $\gamma=1$ at $z=0.6$.
In the old solution with constant dark energy
\begin{align}
{H^2(z)\over H_0^2}=
0.8+0.3(1+z)^3+10^{-4}(1+z)^4.\end{align}
$z=0.6$ corresponds to $\gamma=0.9$.
The most promising  way to find time dependence of dark energy still is to measure the equation of states \cite{KKM} \cite{Planck18}. 

\begin{figure}
\begin{center}
\includegraphics[height=23pc]{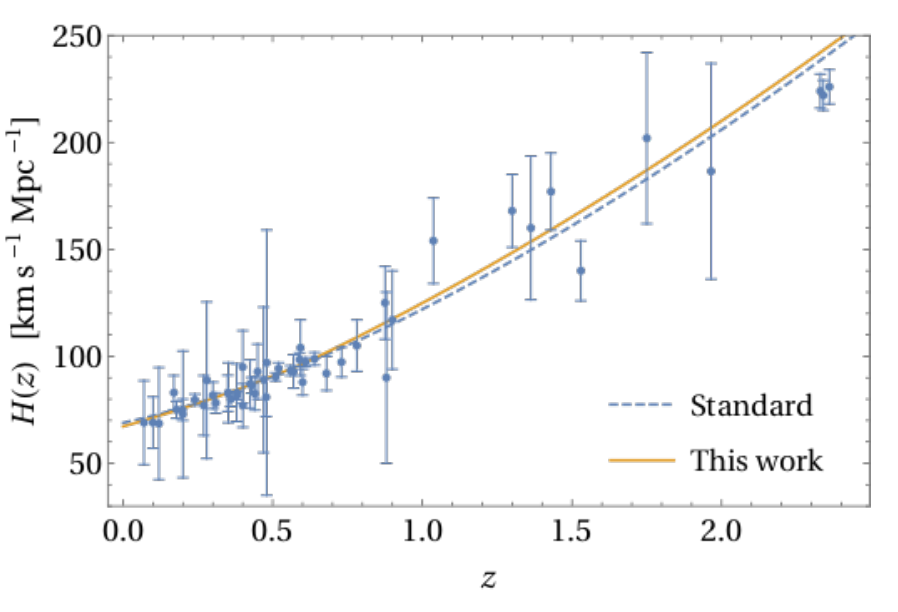}
\caption{\label{fig.Hubble}
The Hubble parameter measurements \cite{GZ} and their errors (in units of ${km/ s Mpc}$) are compared with theoretical predictions.}
\end{center}
\end{figure}

Before concluding this section, we remark on the Baryo-genesis.
The mass term of the right-handed neutrino is
\begin{align}
\int d^4x \omega(t) M_R^{ab}\psi_a\psi_b,
\label{RNMS}
\end{align}
where the indices $a,b$ denote 3 generations.
They could be responsible for Baryo-genesis through lepto-genesis \cite{FY}.
As is well known, one of the necessary condition for lepto-genesis  is non-equilibrium physics.
In other words, we may assume $2M_R< H$.
The observed baryon to photon ratio $n_B/s\sim 10^{-10}$ could be explained by the following parameters \cite{AHKY}.
\begin{align}
m_{\nu3}\sim 0.05eV,~~\delta_{CP}\sim 1,~~
H>2 M_R\sim 10^8GeV.
\end{align}
So lepto-genesis takes place during reheating process.
We assume $H > 2M_R$ at the end of the inflation. $H$ will decrease further down to $10^6 GeV$
by reheating.
The heavy neutrino with $M_R \sim 10^8GeV$ may be pair produced  when $H>2M_R$.   Lepto-genesis is accomplished in such a non-equilibrium process.

It is amusing to note that many  small parameters are of the same order in our scenario.
 \begin{align}
 P\sim n_B/s\sim \text{Dark energy scale/Hadron scale} \sim10^{-10}.
 \end{align}
 Since $M_R$ decays immediately after the pair creation, 
 it fixes the initial conditions of the neutrinos.  
They will not be pair produced after the temperature drops $T<2M_R$.
 
\section{Conclusions and Discussions}
\setcounter{equation}{0}
In the concluding section, we would like to reflect why the  cosmological constant
problem was thought to be one of the hardest problem and which aspect is vulnerable.
It is correct that a large amount of Energy has to be dispersed to reduce the cosmological constant.
In order to make a room for the dark energy, we need a mechanism for almost complete entropy transfer.
The surprising feature of dark energy is its short life time. 
In fact it decreased fastest among radiation and matter. So the reducing cosmological constant
took place literally without noticing by any of us until QCD phase transition.
The engine of the self destruction has been QCD anomaly and curvature perturbation.
Dark energy creates a pair of gluons and photons. The dark energy turns into quarks and leptons by their weak interaction decay chains.
The nature of dark energy drastically changes at the QCD phase transition.

When the Universe cools down around $100 MeV$, the quarks and gluons bound into hadrons.
The dark energy does not have enough energy to create hadrons anymore. So at this point
stable or mature dark energy is born. Since the life time of dark energy is governed by
pair photon creation and it is as long as the life time of the Universe, dark energy becomes the most stable form of energy akin to cosmological constant.
 
We can adopt the experimental value as the curvature perturbation to investigate a
history of dark energy. At the  big bang, thermalized radiation dominates and dark energy
density dwindles from the initial $50\%$. Dark  energy continues to produce entropy by decaying with gluon pair production. However it has become stable after QCD transition since it lost major decay channel.
Nevertheless, dark energy reappears as the dominant component of the energy densities of the Universe. Finally it becomes such dominant that the Universe resumed accelerated expansion.
We have thus recreated a history of dark energy which can explain why it dominates now.
The present dark energy
is a self consistent solution of Friedmann's equation with Anomaly. Our Universe after all keeps the subtle balance between the classical and quantum physics. The observations of curvature perturbation are expected to be accurate on dark energy scale. On the other hand, we need to have much better data on those coming out the Horizon much earlier.

The most unexpected element is the curvature perturbation who supplies energy  for many reheating processes.  Stable supply of energy has been crucial even for dark energy. The successful strategy is to require the  Hubble 
scale of dark energy is given by decay width of inflaton or conformal mode.  $P\alpha_e^3T_{DE}\sim H$ . The hierarchy $H/M_P\sim 10^{-60}$
is reproduced. This is due to the small QED coupling $\alpha_e^3$ and small curvature perturbation. $P\sim 10^{-9}$.
We have identified  all  suppression factors to resolve  the cosmological constant problem. 

However it is clear we need small $P\sim 10^{-9}$ to begin with. 
Although we have found a clue on the notorious $10^{-60}$ factor,
we encounter the small hierarchy problem of why curvature perturbation $10^{-9}$ is so small .
This problem may be  solved by finding a minimum of $H(\omega )$ in such a way that $H/M_P \sim 10^{-4}$.
We find it encouraging that power potentials can be obtained by solving renormalization group
equation \cite{YK}.
It is reassuring that  our Universe satisfies general relativity and quantum field theory.
UV-IR mixing is necessary for the existence of dark energy.
The short distance effect does not spoil de Sitter invariance. Anomaly gives rise to Casimir effect after
cosmological constant is subtractable. On the other hand, IR effects may be dual to the power potential
and cosmological constant is subtractable again. We have shown that UV-IR mixing effects could give rise to  naturally small dark energy.

The dark energy is the most likely candidate for the cosmological constant now.
It is non vanishing if curvature  perturbations exist just like CMB. Not only that
conformal anomaly in gauge theory is indispensable to explain the magnitude of dark energy.
The both IR and UV effects are necessary unless it can be subtracted.
The magnitude of dark energy  can be explained naturally in this approach.
The necessary condition is that we engineer time dependent quantum gravity.
We have fixed the theory to be general covariant and gauge invariant. Nevertheless we can 
investigate time dependent phenomena such as dark energy. 
It is conceivable that fundamental law itself evolves from the humble origin. Such a possibility
is also interesting to explore\cite{Novikov}.

\section*{Acknowledgment}
This work is supported by Grant-in-Aid for Scientific Research (C) No. 16K05336. I thank Chong-Sun Chu, Satoshi Iso, Jun Nishimura, Hikaru Kawai, Kazunori Kohri, Takahiko Matsubara ,Hirotaka Sugawara 
and especially Hiroyuki Kitamoto for discussions.


\begin{thebibliography}{99}
 
 \bibitem{WS}
 Steven Weinberg
 Rev.Mod.Phys. 61 (1989) 1-23
  \bibitem{Polyakov}
A. M. Polyakov, 
Nucl. Phys. B \textbf{797}, 199 (2008) 
[arXiv:0709.2899 [hep-th]]. 
\bibitem{Jackiw}
R. Jackiw, C. Nunez and S.-Y. Pi, 
Phys. Lett. A \textbf{347}, 47 (2005)
[hep-th/0502215].
\bibitem{TW}
N. C. Tsamis and R. P. Woodard,
Nucl. Phys. B \textbf{474}, 235 (1996)
[hep-ph/9602315].
\bibitem{KKN}
H. Kawai, Y. Kitazawa and M. Ninomiya,
Nucl. Phys. B \textbf{404}, 684 (1993)
[hep-th/9303123].
\bibitem{InfAG}
A. H. Guth, \textit{The Inflationary Universe} (Perseus Books, New York, 1997) 
\bibitem{ksato}
K. Sato, Monthly Notices of Royal Astronomical Society, 195, 467, (1981).
\bibitem{InfLD}
A. D. Linde,
Phys. Lett. \textbf{108B}, 389 (1982)
[Adv. Ser. Astrophys. Cosmol. \textbf{3}, 149 (1987)].
\bibitem{InfPS}
A. Albrecht and P. J. Steinhardt,
Phys. Rev. Lett. \textbf{48}, 1220 (1982)
[Adv. Ser. Astrophys. Cosmol. \textbf{3}, 158 (1987)]. 
\bibitem{KKM}
H. Kitamoto and Y. Kitazawa, T.  Matsubara
Phys.Rev.D 101 (2020) 2, 023504 [arXiv: 1908.02534 [hep-th]]
\bibitem{YK}
Y. Kitazawa,
 [arXiv: 2011.14640 [hep-th]]

\bibitem{RP}
B. Ratra and P. J. E. Peebles, 
Phys. Rev. D \textbf{37}, 3406 (1988). 
\bibitem{CDS}
R. R. Caldwell, R. Dave and P. J. Steinhardt,
Phys. Rev. Lett. \textbf{80}, 1582 (1998)
[astro-ph/9708069].j)
\bibitem{Infst}
A. A. Starobinsky,
JETP Lett. \textbf{30}, 682 (1979)
[Pisma Zh. Eksp. Teor. Fiz. \textbf{30}, 719 (1979)]
\bibitem{GW}
D. J. Gross and F. Wilczek, Phys. Rev. Lett. 30, 1343 (1973).
\bibitem{HP}
H. D. Politzer, Phys. Rev. Lett. 30, 1346 (1973).

\bibitem{LL}
R. Liddle and S. M. Leach,
Phys. Rev. D 68, 103503


\bibitem{Starobinsky1985}
A. A. Starobinsky, Phys. Lett. B117, 175 (1982); JETP Lett. 42, 152 (1985).

 \bibitem{Planck18}
     N. Aghanim et al. [Planck Colaboration],
     [arXiv:1807.06209[astro-ph.CO].
 
 \bibitem{ABYRM}
Rouzbeh Allahverdi, Robert Brandenberger, Francis-Yan Cyr-Racine, Anupam Mazumdar,
 [arXiv:1001.2600 [hep-th]].

\bibitem{GZ}
J. Maga\~na, M. H. Amante, M. A. Garcia-Aspeitia and V. Motta,
Mon. Not. Roy. Astron. Soc. \textbf{476}, no. 1, 1036 (2018),
[arXiv:1706.09848 [astro-ph.CO]].

\bibitem{Bond}
 D. S. Salopek and J. R. Bond, Phys. Rev. D42, 3936 (1990).

\bibitem{Stewart}
M. Sasaki and E. D. Stewart, Prog. Theor. Phys. 95, 71 (1996), arXiv:astro-ph/9507001.

\bibitem{Lyth2005a}
 D. H. Lyth, K. A. Malik, and M. Sasaki, JCAP 0505, 004 (2005), arXiv:astro-ph/0411220v3.

\bibitem{Lyth2005b}
 D. H. Lyth and Y. Rodriguez,
Phys.Rev.Lett. 95, 121302 (2005),
arXiv:astro-ph/0504045. 

\bibitem{FY}
M.Fukugita, T. Yanagida, 
Phys.Lett. B174 (1986) 45.

\bibitem{AHKY}
T. Asaka, K. Hamaguchi, M. Kawasaki, T. Yanagida, 
Phys.Lett. B464(1999) 12,
arXiv:hep-ph/9906366.

\bibitem{Novikov}
E.  Novikov,
Amer.Res. J. Physics, v.4(1),1-9, (2018).
  
\end{thebibliography}
\end{document}